\documentclass[epj]{svjour}

\usepackage{graphicx}
\usepackage{amsmath}
\usepackage{amsbsy}

\newcommand{\bk}{{\bf k}}

\begin{document}

\title{Effect of strain-induced electronic topological transitions on
   the superconducting properties of La$_{2-x}$Sr$_x$CuO$_4$ thin
   films}

\titlerunning{Strain-induced ETT in LSCO thin films}

\author{G. G. N. Angilella\inst{1} 
   \and G. Balestrino\inst{2,5} 
   \and P. Cermelli\inst{3}
   \and P. Podio-Guidugli\inst{4}
   \and A. A. Varlamov\inst{5}}

\institute{Dipartimento di Fisica e Astronomia, Universit\`a di Catania,
and Istituto Nazionale per la Fisica della Materia, UdR di Catania,
Corso Italia, 57, I-95129 Catania, Italy
\and
Dipartimento di Scienze e Tecnologie Fisiche ed
   Energetiche,
Universit\`a di Roma ``Tor Vergata'', Via di Tor Vergata, 110,
   I-00133 Roma, Italy
\and
Dipartimento di Matematica, Universit\`a di Torino, Via Carlo Alberto, 
   10, I-10123 Torino, Italy
\and
Dipartimento di Ingegneria Civile,
Universit\`a di Roma ``Tor Vergata'', Via di Tor Vergata, 110,
   I-00133 Roma, Italy
\and
Istituto Nazionale per la Fisica della Materia, UdR di Tor Vergata,
Via di Tor Vergata, 110, I-00133 Roma, Italy}

\date{\today}

\abstract{%
We propose a Ginzburg-Landau phenomenological model 
   for the dependence of the critical temperature on microscopic
   strain in tetragonal high-$T_c$ cuprates.
Such a model is in agreement with the experimental results for LSCO
   under epitaxial strain, as well as with the hydrostatic pressure
   dependence of $T_c$ in most cuprates.
In particular, a nonmonotonic dependence of $T_c$ on hydrostatic
   pressure, as well as on in-plane or apical microstrain, is derived.
From a microscopic point of view, such results can be understood as
   due to the proximity to an electronic
   topological transition (ETT).
In the case of LSCO, we argue that such an ETT can be driven by a
   strain-induced modification of the band structure, at constant hole
   content, at variance with a doping-induced ETT, as is usually
   assumed.
\PACS{
{74.62.Fj}{Transition temperature variations; pressure effects}
\and
{74.20.De}{Phenomenological theories (two-fluid, Ginzburg-Landau, etc.)}
\and
{74.72.Dn}{La-based cuprates}
     } 
} 

\maketitle

\section{Introduction}

The application of high pressure to high-$T_c$ cuprate superconductors 
   (HTS) is known to modify remarkably the superconducting properties
   of these materials \cite{Wijngaarden:99}.
In particular, it was shown by Gao \emph{et al.} \cite{Gao:94} that
   quasi-hydrostatic pressure can increase the critical temperature of 
   Hg\-Ba$_2$\-Ca$_2$\-Cu$_3$\-O$_{8+ \delta}$ up to about 164~K. 
Such a record has been very recently challenged, by achieving $T_c = 117$~K 
   in C$_{60}$ single crystals, where an expansion of the crystalline
   lattice was realized via the intercalation of CHCl$_3$ and
   CHBr$_3$ (chemical pressure) \cite{Schoen:01}.
Even more promising is the possibility of increasing $T_c$
   in HTS thin films via the anisotropic strain induced by
   epitaxial growth on mismatching substrates. 
The effect of tensile and compressive
   epitaxial strains on the transport properties has been investigated. 
This research has been mostly focused on
   the La$_{2-x}$Sr$_x$CuO$_4$ (LSCO) compound because in this
   system the hole concentration is well controlled over an
   exceptionally wide range, and mostly determined by the Sr content
   (together with small oxygen non-stoichiometry).
Using SrLaAlO$_4$ (SLAO) substrates (with in-plane lattice spacing
   $a=3.755$~\AA), epitaxial LSCO films ($a=3.777$~\AA) have been grown
   which were in-plane compressively strained \cite{Trofimov:94}.
Critical temperatures as high as 49~K have been obtained 
   in slightly underdoped La$_{2-x}$Sr$_x$CuO$_4$ with $x=0.11$
   \cite{Locquet:98}, and $T_c = 44$~K in the same compound, at
   optimal doping ($x=0.15$) \cite{Sato:97}.
Recently, it has been shown that a compressive epitaxial strain can
   induce an insulator-superconductor transition in undoped or slightly
   doped La$_2$CuO$_4$ films \cite{Si:01}.
Such an epitaxial-strain-induced transition is a further dramatic
   demonstration of the significance of strain in HTS materials. 

Different mechanisms have been proposed to explain the dependence of
   $T_c$ on lattice strain \cite{Sato:00,Chen:00,Bianconi:00}.  
A simple explanation is based on the possible dependence on strain of
   the oxygen excess in the LSCO structure. 
However, it has been shown in Ref.~\cite{Si:99} that this cannot be
   the only explanation. 
There is now a general agreement that a key to understanding the
   relationship between epitaxial strain and superconducting
   properties is the microstrain associated to certain parameters
   describing the fine structure of the LSCO cell. 
Locquet \emph{et al.} \cite{Locquet:98} have recently suggested that
   the most relevant microparameter is the distance between the Cu
   ions in the CuO$_2$ planes and the apical oxygen. 

A phenomenological model accounting for the role of the apical
   distance has been developed in Ref.~\cite{Cermelli:01}; this model,
   which accounts also for other microparameters, is based on a
   Ginzburg-Landau approach. 
Needless to say, no phenomenological model could by itself clarify the
   physical mechanism connecting microstrains and superconducting
   properties. 
Yet, in our opinion, a result of Ref.~\cite{Cermelli:01} offers a clue
   for a possible explanation: we refer to the prediction, compatible
   with the available experimental data for LSCO, that the critical
   temperature is a nonlinear function of the epitaxial strain
   $\varepsilon_{\mathrm{epi}}$ in the experimentally accessible range
   $-0.006\leq\varepsilon_{\mathrm{epi}}\leq 0.006$. 
This prediction is directly reminiscent, for reasons that we now
   detail, of a general prediction of the theory of electronic
   topological transitions (ETT)
   \cite{Varlamov:89,Blanter:94,Markiewicz:97}.

The effect of an ETT on the superconducting properties of quasi-2D systems,
   such as HTS materials, has attracted renewed interest
   \cite{Onufrieva:99a,Onufrieva:99b}.
It is known that optimally-doped HTS materials are in the proximity of
   an ETT from a hole-like to an electron-like Fermi surface; in the
   case of LSCO, angle-resolved photoemission spectroscopy (ARPES) has
   shown that such a transition occurs for $x\simeq0.2$ (slightly
   overdoped samples) \cite{Ino:01}.
Now, an ETT can be driven, in addition to doping, by a number of
   different external agents, such as impurity concentration,
   hydrostatic pressure and, as we here surmise, anisotropic strain. 
In Ref.~\cite{Angilella:01,Angilella:02}, the dependence of the
   critical temperature $T_c$ on the parameter $z$ measuring the
   deviation of the chemical potential from the ETT was studied, and
   found to be nonmonotonic.
In this paper, we propose that the physical mechanism at the origin of
   the change in the critical temperature in LSCO films under epitaxial strain
   is an ETT, driven by microstructural deformations. 
Furthermore, we show that the nonmonotonic behavior, predicted in the
   ETT scenario, is in agreement with the behavior foreseen by the
   phenomenological model of Ref.~\cite{Cermelli:01}.

The outline of the paper is as follows.
After a brief review of the Ginzburg-Landau phenomenological model of
   Ref.~\cite{Cermelli:01}, relating $T_c$ to the epitaxial strain
   $\varepsilon_{\mathrm{epi}}$ (Sec.~\ref{sec:phenomenological}), we
   introduce a generic microscopic model for a superconducting
   electron system on a square lattice, close to an ETT
   (Sec.~\ref{sec:microscopic}).
Our numerical results are presented in Sec.~\ref{sec:numerical}, where a
   nonmonotonic dependence for $T_c$ as a function of hole doping and band
   structure is recognized, in agreement with the
   phenomenological model of Sec.~\ref{sec:phenomenological}.
Conclusions and directions for future work are the subject of
   Sec.~\ref{sec:conclusions}.

\section{Phenomenological model}
\label{sec:phenomenological}

Experimental data show that, in high-$T_c$ materials such as YBCO and
   LSCO, the critical temperature has a parabolic dependence on
   applied hydrostatic pressure: as pressure increases, so does $T_c$
   until it reaches a maximum, after which it decreases
   \cite{Wijngaarden:99}.
Different trends have also been recorded (notably, in orthorhombic
   YBCO) for $T_c$ as a function of uniaxial strain
   \cite{Welp:92,Budko:92}.
These have 
   been interpreted as evidence for the importance of the internal
   strains, especially in non-tetragonal compounds \cite{Pickett:97}.
Moreover, the role of oxygen relaxation processes in establishing
   hysteresis loops in the pressure-temperature history of YBCO has
   been emphasized \cite{Tissen:99}.
On the other hand, such subtleties in the pressure dependence of $T_c$ 
   can be neglected in tetragonal LSCO \cite{Yamada:92}, whose hole
   content is mainly determined by the amount of doping Sr.
Therefore, LSCO in the tetragonal phase is an ideal
   candidate to study 
   the dependence of $T_c$ on applied pressure, without (much of) the
   complication arising from pressure-induced charge rearrangements.

Here, we show that a nonmonotonic dependence of $T_c$ on applied
   pressure is predicted by a modified Ginzburg-Landau model, which
   takes into account the dependence of $T_c$ on the lengths of the
   apical and planar Cu--O bonds.
This model has been first introduced for application to LSCO films
   under epitaxial strain \cite{Cermelli:01}.
Remarkably, the predicted behavior of these films is similar: for an
   increasing, \emph{compressive} epitaxial strain, the critical
   temperature rises to a maximum, then it decreases.

Tetragonal LSCO has a perovskite lattice structure, with the Cu atoms
   in octahedral coordination with the O atoms.
We denote by (Cu--O)$^a$ and (Cu--O)$^b$ the Cu--O distances in the
   $ab$ plane (\emph{viz.,} the half-diagonals of the Cu--O octahedron 
   in that atomic plane), and by (Cu--O)$^{\mathrm{api}}$ the apical
   distance (\emph{viz.,} the half-diagonal of the Cu--O octahedron in 
   the direction of the $c$ axis).
We also introduce the \emph{microscopic strain measures}
\begin{subequations}
\begin{eqnarray}
p_a &:=& \frac{{\mbox{(Cu--O)}}^a -
   {\mbox{(Cu--O)}}^a_0}{{\mbox{(Cu--O)}}^a_0} ,\\
p_b &:=& \frac{{\mbox{(Cu--O)}}^b -
   {\mbox{(Cu--O)}}^b_0}{{\mbox{(Cu--O)}}^b_0} ,\\
p_{\mathrm{api}} &:=& \frac{{\mbox{(Cu--O)}}^{\mathrm{api}} -
   {\mbox{(Cu--O)}}^{\mathrm{api}}_0}{{\mbox{(Cu--O)}}^{\mathrm{api}}_0} ,
\end{eqnarray}
\end{subequations}
where ${\mbox{(Cu--O)}}^a_0$, ${\mbox{(Cu--O)}}^b_0$, and
   ${\mbox{(Cu--O)}}^{\mathrm{api}}_0$ are reference values for the
   corresponding interatomic distances.
We let $\boldsymbol{\varepsilon}$ denote the \emph{macroscopic strain
   tensor,} with components $\varepsilon_{aa}$, $\varepsilon_{bb}$,
   etc.: $\boldsymbol{\varepsilon}$ measures the overall strain of the 
   unit cell, whereas the microscopic strains measure relative changes 
   in the interatomic distances within the cell.

For $\varphi$ the superelectron density, we write the Ginzburg-Landau
   free energy as
\begin{equation}
G = G_0 (T,\boldsymbol{\varepsilon}) + a_0 \alpha(T,p_a ,p_b
   ,p_{\mathrm{api}} ) \varphi^2 + b_0 \varphi^4 ,
\end{equation}
where $a_0, b_0 >0$ are constants, and the function $\alpha$ accounts
   for the dependence of the critical temperature on the Cu--O
   distances.
To fix the ideas, we assume a quadratic dependence, in the form
\begin{eqnarray}
\alpha(T,p_a ,p_b ,p_{\mathrm{api}} ) &=& T - T_c^0 -\lambda_1 (p_a +
   p_b ) - \mu_1 (p_a^2 + p_b^2 ) \nonumber\\
&& -\lambda_2 p_{\mathrm{api}} -\mu_2 p_{\mathrm{api}}^2 
-\sigma (p_a + p_b ) p_{\mathrm{api}} .
\end{eqnarray}
As is well known, the vanishing of $\alpha$ determines the critical
   temperature:
\begin{eqnarray}
T_c &=& T_c^0 +\lambda_1 (p_a +
   p_b ) + \mu_1 (p_a^2 + p_b^2 ) \nonumber\\
&&+\lambda_2 p_{\mathrm{api}} +\mu_2
   p_{\mathrm{api}}^2 +\sigma (p_a + p_b ) p_{\mathrm{api}} .
\label{eq:TcGL}
\end{eqnarray}
We estimate the phenomenological coefficients $\lambda_i$, $\mu_i$ and 
   $\sigma$ from the available experimental data for LSCO under
   strain.
Our procedure consists of two steps: (i) we determine how the
   microstrains $(p_a , p_b , p_{\mathrm{api}} )$ depend on the
   macroscopic strain $\boldsymbol{\varepsilon}$; (ii) we express the
   critical temperature in Eq.~(\ref{eq:TcGL}) as a function of the
   strain $\boldsymbol{\varepsilon}$ and fit the resulting expression
   to the available data on strain and $T_c$.

Step (i) has been performed in Ref.~\cite{Cermelli:01}, on the basis
   of the experimental findings of Locquet \emph{et al.}
   \cite{Locquet:98a} for the variation of the interatomic distances
   in epitaxially strained thin films of LSCO.
In the tetragonal phase, it is reasonable to assume that $p_a =
   \varepsilon_{aa}$, $p_b = \varepsilon_{bb}$, but the experimental
   data in Ref.~\cite{Locquet:98a} show that $p_{\mathrm{api}}$ is a
   highly nonlinear function of the principal strains
   $(\varepsilon_{aa} , \varepsilon_{bb} , \varepsilon_{cc} )$.
The actual analytical expression for $p_{\mathrm{api}} =
   \tilde{p}_{\mathrm{api}} (\varepsilon_{aa} , \varepsilon_{bb} ,
   \varepsilon_{cc} )$, interpolating the data in
   Ref.~\cite{Locquet:98a}, has been determined in
   Ref.~\cite{Cermelli:01}.

Step (ii) corresponds to substituting into Eq.~(\ref{eq:TcGL}) the
   expressions for $(p_a , p_b , p_{\mathrm{api}} )$ in terms of
   $(\varepsilon_{aa} , \varepsilon_{bb} , \varepsilon_{cc} )$.
The result is an expression of the form
\begin{eqnarray}
T_c &=& \tilde{T}_c (\varepsilon_{aa} , \varepsilon_{bb} ,
   \varepsilon_{cc} ) \nonumber\\
&=& T_c^0 + \lambda_1 (\varepsilon_{aa} + \varepsilon_{bb} ) + \mu_1
   (\varepsilon_{aa}^2 + \varepsilon_{bb}^2 ) \nonumber\\
&& + \lambda_2
   \tilde{p}_{\mathrm{api}} (\varepsilon_{aa} , \varepsilon_{bb} ,
   \varepsilon_{cc} )
+ \mu_2 \tilde{p}_{\mathrm{api}}^2 (\varepsilon_{aa} ,
   \varepsilon_{bb} , \varepsilon_{cc} ) \nonumber\\
&&+ \sigma (\varepsilon_{aa} +
   \varepsilon_{bb} ) \tilde{p}_{\mathrm{api}} (\varepsilon_{aa} ,
   \varepsilon_{bb} , \varepsilon_{cc} ),
\label{eq:Cerm}
\end{eqnarray}
which still contains the unknown coefficients $\lambda_i$, $\mu_i$,
   and $\sigma$.
These parameters may be determined by fitting the expression
   (\ref{eq:Cerm}) to the experimental data on the dependence of the
   critical temperature under strain \cite{Cermelli:01}.
Using the data in
   Ref.~\cite{Locquet:98,Welp:92,Locquet:98a,Budko:92,Chen:91}, we
   obtain the values listed in Table~\ref{tab:coeff}.

\begin{table}
\centering
\begin{tabular}{cr@{.}lr@{.}lc}
\hline
\hline
 & \multicolumn{2}{c}{$\lambda_i \times 10^{-3}$}
 & \multicolumn{2}{c}{$\mu_i \times 10^{-3}$} 
 & $\sigma \times 10^{-3}$ \\
\hline
$i=1$ & $3$ & $962$ & $-579$ & $664$ & $5$ \\
$i=2$ & $5$ & $028$ & $-42$  & $029$ & $-$ \\
\hline
\hline
\end{tabular}
\caption{Calculated coefficients (in K) for the dependence of $T_c$ on 
   the changes of dimensions of the CuO octahedron,
   Eq.~(\protect\ref{eq:TcGL}).} 
\label{tab:coeff}
\end{table}

\begin{remark}
At first order in $(p_a , p_b , p_{\mathrm{api}} )$, only the linear
   terms are important, and the expression (\ref{eq:TcGL}) reduces to
\begin{equation}
T_c \sim T_c^0 + [4(p_a + p_b ) + 5 p_{\mathrm{api}} ] \times 10^3 ,
\end{equation}
which shows that the critical temperature increases with the size of
   the Cu--O octahedron, and is nearly isotropic in the horizontal and 
   vertical microstrains.
\end{remark}

\begin{figure}
\centering
\includegraphics[bb=80 80 534 715,clip,height=0.9\columnwidth,angle=-90]{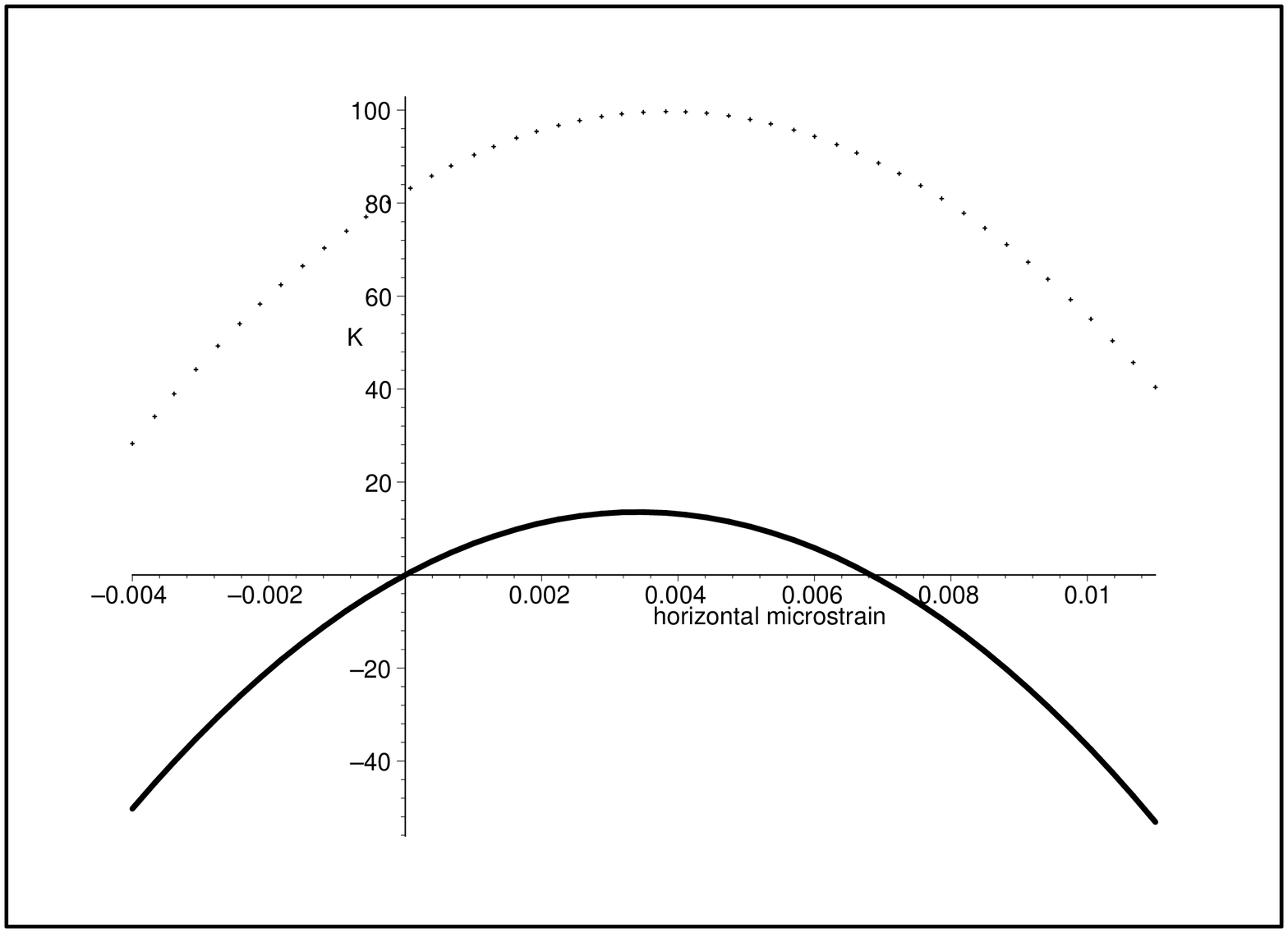}
\caption{Theoretical variation of the critical temperature $T_c -
   T_c^0$ (in K) on the planar microstrain $p_a$, for two fixed values 
   of the apical microstrain, $p_{\mathrm{api}} = 0$ (solid line), and 
   $p_{\mathrm{api}} = 0.1$ (dotted line), according to
   Eq.~(\protect\ref{eq:TcGL}).}
\label{fig:Tcpa}
\end{figure}

\begin{figure}
\centering
\includegraphics[bb=80 80 534 715,clip,height=0.9\columnwidth,angle=-90]{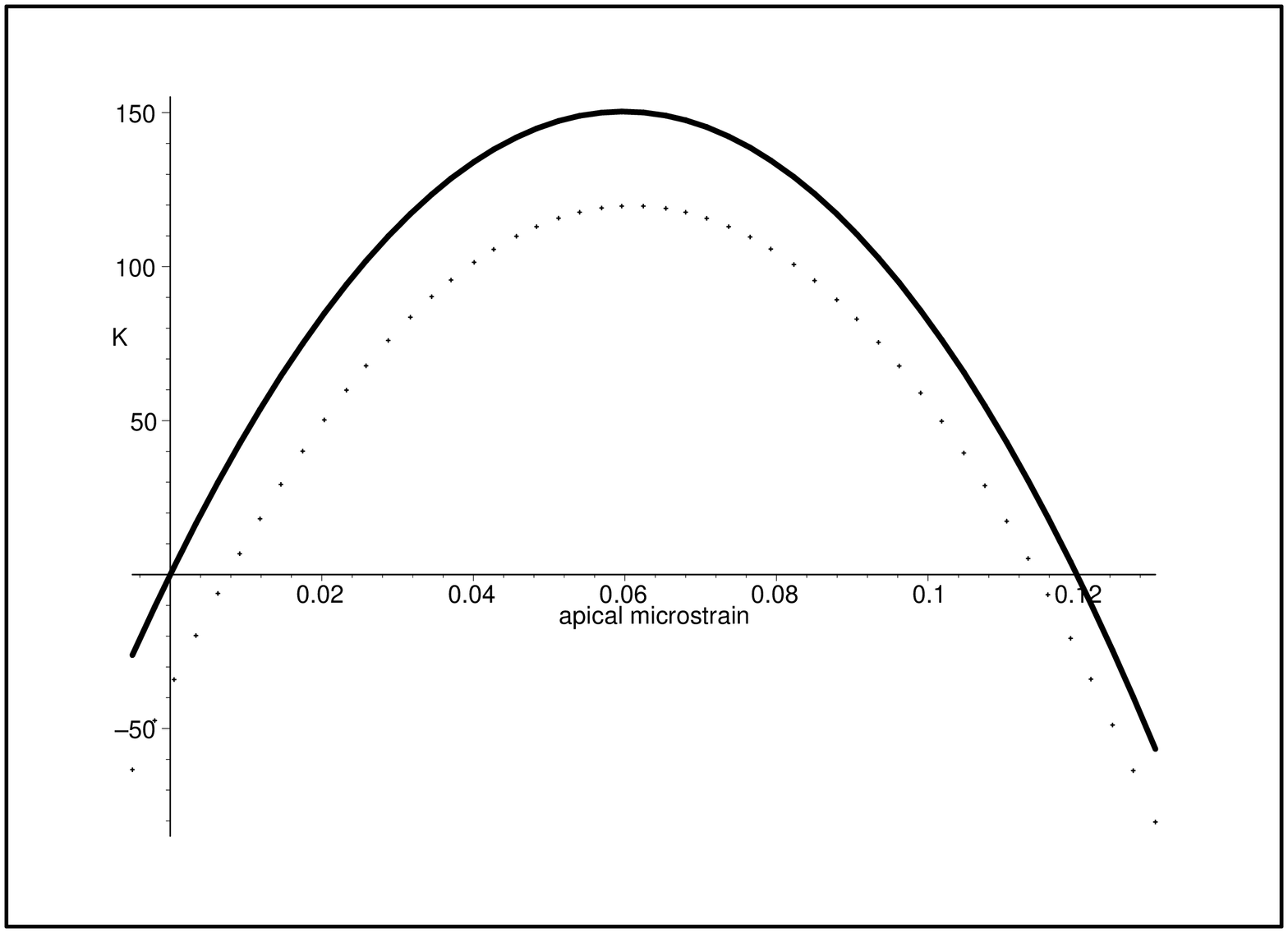}
\caption{Theoretical variation of the critical temperature $T_c -
   T_c^0$ (in K) on the apical microstrain $p_{\mathrm{api}}$, for two
   fixed values  
   of the planar microstrain, $p_a = 0$ (solid line), and 
   $p_a = 0.01$ (dotted line), according to
   Eq.~(\protect\ref{eq:TcGL}).}
\label{fig:Tcpapi}
\end{figure}

\begin{remark}
More importantly, for $p_{\mathrm{api}}$ fixed, the critical
   temperature reaches a maximum in correspondence of a given
   horizontal microstrain $p_a = p_b = p_a^{\mathrm{max}}$, and then
   it starts decreasing (Fig.~\ref{fig:Tcpa}).
A completely analogous behavior takes place for $p_a$ and $p_b$ fixed: 
   the critical temperature reaches a maximum at a given apical
   microstrain $p_{\mathrm{api}} = p_{\mathrm{api}}^{\mathrm{max}}$
   (Fig.~\ref{fig:Tcpapi}).
\end{remark}

\begin{figure}
\centering
\includegraphics[bb=80 80 534 715,clip,height=0.9\columnwidth,angle=-90]{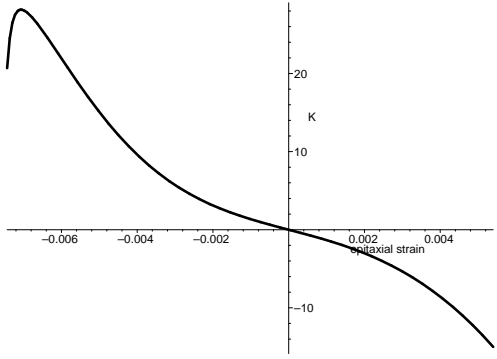}
\caption{Theoretical variation of the critical temperature $T_c -
   T_c^0$ (in K) on the epitaxial strain, according to
   Eq.~(\protect\ref{eq:Cerm}).}
\label{fig:Tcstrain}
\end{figure}

\begin{remark}
Under epitaxial strain, $\varepsilon_{aa} = \varepsilon_{bb} =
   \varepsilon_{\mathrm{epi}}$, and $\varepsilon_{cc} = -
   2( c_{13} / c_{33}) \varepsilon_{\mathrm{epi}}$ ($c_{13}$ and $c_{33}$
   are components of the constant elasticity tensor of the film
   \cite{Cermelli:01}), so that the critical temperature in
   (\ref{eq:Cerm}) is a function of the epitaxial strain
   $\varepsilon_{\mathrm{epi}}$ only.
This function is plotted in Fig.~\ref{fig:Tcstrain}.
Note that $T_c$ is monotonically decreasing in the experimentally
   accessible range $-0.006 \leq \varepsilon_{\mathrm{epi}} \leq
   0.006$, but shows a sharp maximum just below the lower bound of
   this interval: the predicted values of the maximum $T_c$ are not
   very far from the experimentally accessible interval, where the
   quadratic approximation in (\ref{eq:TcGL}) may still be expected to 
   hold.
Thus, the prediction of the phenomenological model seems reasonable.
\end{remark}

\begin{figure}
\centering
\includegraphics[bb=80 80 534 715,clip,height=0.9\columnwidth,angle=-90]{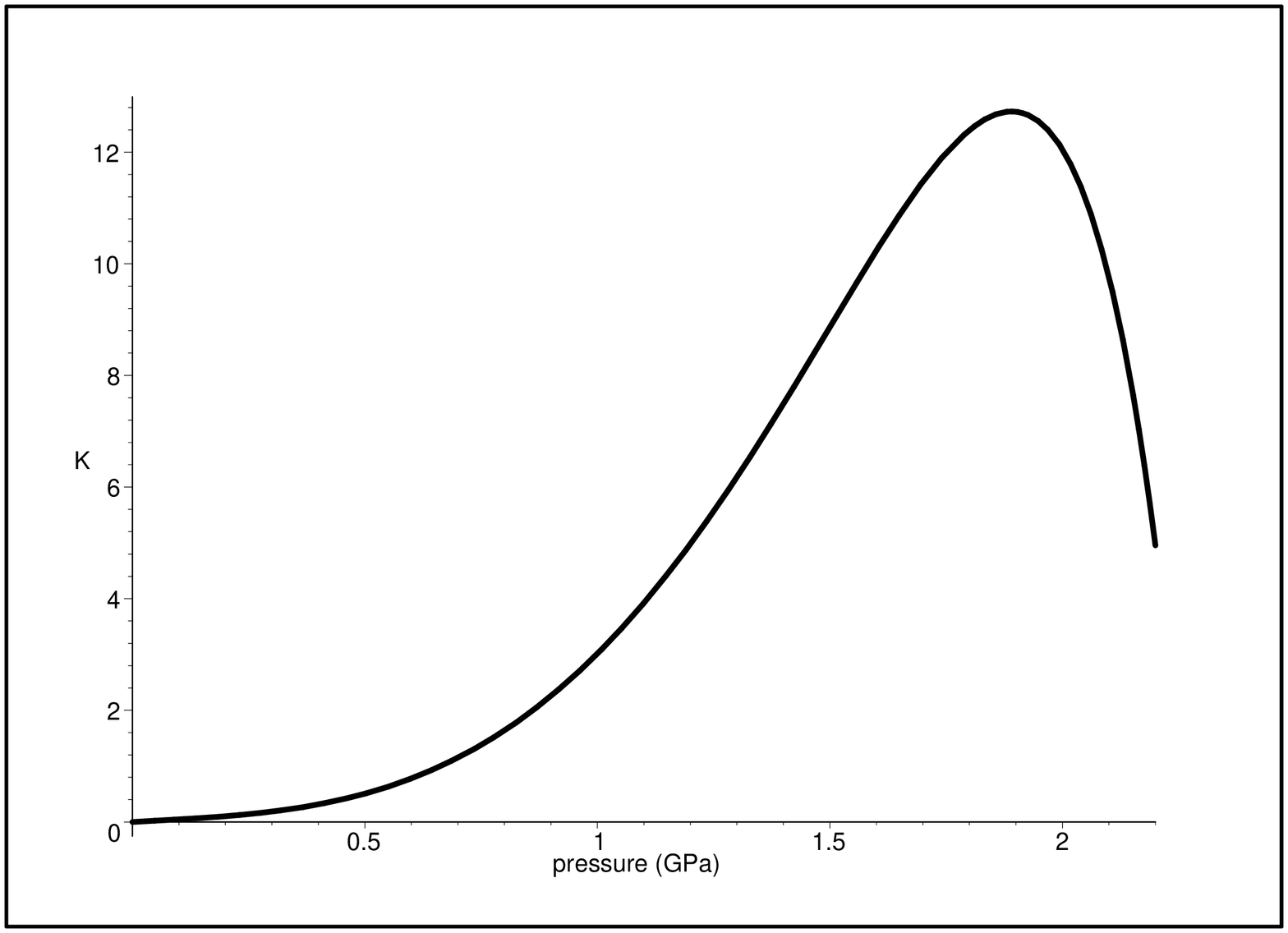}
\caption{Theoretical variation of the critical temperature $T_c -
   T_c^0$ (in K) on applied hydrostatic pressure, according to
   Eq.~(\protect\ref{eq:Cerm}).}
\label{fig:Tchydro}
\end{figure}

\begin{remark}
Under an applied hydrostatic pressure $P$, we have
\begin{subequations}
\begin{eqnarray}
\varepsilon_{aa} = \varepsilon_{bb} &=& \frac{(-c_{33} + c_{13}
   )P}{c_{11} c_{33} -2 c_{13}^2 + c_{12} c_{33}} ,\\
\varepsilon_{cc} &=& - \frac{(c_{11} -2c_{13} + c_{12} )P}{c_{11}
   c_{33} -2c_{13}^2 + c_{12} c_{33}} ,
\end{eqnarray}
\end{subequations}
and the remaining strain components vanish.
The critical temperature in (\ref{eq:Cerm}) becomes a function of
   the applied pressure.
This function is plotted in Fig.~\ref{fig:Tchydro}, which displays the 
   characteristic maximum of $T_c$ versus pressure, as observed
   experimentally \cite{Wijngaarden:99}.
Note that the numerical agreement with experimental data in this case
   is poorer than for epitaxial strain.
\end{remark}

\begin{remark}
For small strains, an approximate phenomenological expression for
   $T_c$ as a function of the applied strain, widely used in the
   literature, is
\begin{equation}
T_c = T_c^0 + A (\varepsilon_{aa} + \varepsilon_{bb} ) + B
   \varepsilon_{cc} ,
\end{equation}
where, for LSCO, $A=-284$~K and $B=851$~K (notably, $A<0$, $B>0$).
The latter expression may be viewed as a linear approximation to
   (\ref{eq:Cerm}), with $A = \left.\frac{\partial
   T_c}{\partial\varepsilon_{aa}} \right|_0 = \left.\frac{\partial
   T_c}{\partial\varepsilon_{bb}} \right|_0$, and $ B = \left.\frac{\partial
   T_c}{\partial\varepsilon_{cc}} \right|_0$.
\end{remark}

Therefore, the critical temperature is a \emph{decreasing} function of 
   the horizontal macroscopic strain, which seems to be in contradiction
   with the fact that, in Eq.~(\ref{eq:TcGL}), $T_c$ is an
   \emph{increasing} function of the horizontal microstrains $p_a$ and 
   $p_b$ (recall in fact that $p_a = \varepsilon_{aa}$, $p_b =
   \varepsilon_{bb}$).
This may be explained by noting that the apical microstrain
   $p_{\mathrm{api}}$ is a very fast decreasing function of the
   horizontal strain $(\varepsilon_{aa} , \varepsilon_{bb} )$
   \cite{Cermelli:01}, and it prevails quantitatively in
   Eq.~(\ref{eq:TcGL}) over the horizontal microstrains.
Thus, according to Eq.~(\ref{eq:TcGL}), the observed decrease of the
   critical temperature under uniaxial horizontal strain (as measured 
   by the negative $A$) seems to be essentially due to the
   accompanying decrease of the apical distance.

\section{Microscopic model}
\label{sec:microscopic}

We start by considering the following Hubbard-like Hamiltonian for
   an interacting electron system on a 2D square lattice
   \cite{Fehrenbacher:95}
\begin{equation}
H = \sum_{\bk\sigma} \xi_\bk c^\dag_{\bk\sigma} c_{\bk\sigma} + \frac{1}{N}
   \sum_{\bk\bk^\prime} V_{\bk\bk^\prime} c^\dag_{\bk\uparrow}
   c^\dag_{-\bk\downarrow} c_{-\bk^\prime \downarrow} c_{\bk^\prime
   \uparrow} .
\label{eq:hamiltonian}
\end{equation}
Here, $c^\dag_{\bk\sigma}$ ($c_{\bk\sigma}$) is a creation
   (annihilation) operator for an electron state with wavevector $\bk$
   and spin projection 
   $\sigma\in\{\uparrow,\downarrow\}$, $N$ is the number of lattice
   sites, and the sums are restricted to the first Brillouin zone
   (1BZ).
We assume the electron-electron interaction in the separable form
   $V_{\bk\bk^\prime} = \lambda g_\bk g_{\bk^\prime}$, where $g_\bk =
   \frac{1}{2} (\cos k_x - \cos k_y )$ is the lowest-order $d$-wave
   lattice harmonic for a square lattice, and $\lambda$ a
   phenomenological coupling constant ($\lambda<0$).

Detailed band structure calculations \cite{Andersen:95} as well as
   ARPES \cite{Shen:95}
   suggest that a tight-binding approximation for the dispersion
   relation $\xi_\bk$
   of most high-$T_c$ cuprates should retain at least nearest (NN, $t$) and
   next-nearest neighbors (NNN, $t^\prime$) hopping.
We then assume the following rigid band-dispersion relation for LSCO:
\begin{equation}
\xi_\bk = -2t(\cos k_x + \cos k_y ) + 4t^\prime \cos k_x \cos k_y - \mu,
\label{eq:dispersion}
\end{equation}
where $\mu$ denotes the chemical potential, and the components of the
   wavevector $\bk$ are measured in units of the inverse lattice
   spacing.
In order to have a flat minimum in $\xi_\bk$ around the $\Gamma$
   point, as observed experimentally \cite{Shen:95}, the condition
   $0<r<\frac{1}{2}$ must be fulfilled.
A nonzero value of the hopping ratio $r=t^\prime /t$ 
   destroys perfect nesting at $\mu=0$ as well as the electron-hole
   symmetry, and is known to stabilize superconductivity against other 
   possible low-energy instabilities \cite{Alvarez:98}.

As the chemical potential $\mu$ in Eq.~(\ref{eq:dispersion}) varies
   from the bottom, $\varepsilon_\bot = -4t(1-r)$, to the top of the
   band, $\varepsilon_\top = 4t(1+r)$, the Fermi line $\xi_\bk =0$
   evolves from an electron-like contour, closed around the $\Gamma$
   point, to a hole-like contour, whose continuation into higher
   Brillouin zones closes around the $M=(\pi,\pi)$ point (see also
   Fig.~\ref{fig:ETT} below). 
In doing so, an ETT is traversed at $\mu=\varepsilon_c = -4t^\prime$,
   where the Fermi line touches the zone boundaries.

It is worth emphasizing that the assumed $d$-wave momentum dependence
   of the potential energy correlates in a nontrivial way with the
   behaviour of the Fermi line close to the ETT.
Indeed, the above choice for the pairing potential yields a gap energy 
   $\Delta_\bk \propto g_\bk$, with maximum amplitudes occurring at
   $X=(0,\pi)$ (and symmetry related points), \emph{i.e.} exactly at
   the ETT.
Moreover, this is where the shape of the Fermi line is most sensible
   to changes in the hopping ratio $r$ \cite{Hlubina:95}. 
Therefore, a deformation of the Fermi line induces a change of the
   phase space effectively probed by the electron-electron interaction
   \cite{Hodges:71}.
This is particularly relevant in the case of anisotropic pairing with
   $d$-wave symmetry, such as that mediated by the exchange of
   antiferromagnetic spin density wave \cite{Millis:90} or charge
   density wave fluctuations \cite{Perali:96}, as well as in the case
   of $d$-wave pairing enhanced by interlayer pair-tunneling
   \cite{Chakravarty:93,Angilella:99}.

An ETT gives rise to anomalous behaviors in the normal as well as in the 
   superconducting properties of the electron system, as a function of 
   the distance $z=\mu-\varepsilon_c$ from the ETT
   \cite{Varlamov:89,Blanter:94,Markiewicz:97}. 
At variance with the 3D case, a 2D superconductor close to an ETT is
   characterized by a nonmonotonic dependence of $T_c$ on $z$, as
   observed experimentally as a function of doping \cite{Zhang:93}, or
   hydrostatic pressure \cite{Wijngaarden:99}.
Such a result has been recently rederived analytically
   \cite{Angilella:01,Angilella:02}.
Moreover, one finds that $T_c$
   at optimal doping (\emph{i.e.,} near the ETT)
   correlates directly with the hopping ratio $r$, both for an
   $s$- and for a $d$-wave superconductor
   \cite{Angilella:01,Angilella:02}, as is exctracted from band
   structure calculations
   for several hole-doped high-$T_c$ cuprates \cite{Pavarini:01}.

The effects of the proximity to an ETT in the normal state are more
   difficult to be detected.
For example, it is well known that the presence of an ETT at $T=0$
   gives rise to a peak in the thermoelectric power of a metal as well 
   as to minima in the voltage-current characteristic of a tunnel junction
   \cite{Varlamov:89}.
However, the increase of temperature is expected to smear such
   effects.
On the other hand, the sign change of the Hall resistivity
   $R_{\mathrm{H}}$ in the cuprates as a function of doping
   \cite{Tamasaku:94} (see also Ref.~\cite{Locquet:96} for measurements of
   $R_{\mathrm{H}}$ in LSCO thin films) has been related to the
   presence of a Van~Hove singularity in the single-particle spectrum
   of LSCO \cite{Avella:98} (see also Ref.~\cite{Markiewicz:97} for a
   review).
Indeed, it has been shown that the sign of the Hall conductivity
   correlates with $\partial\ln T_c /\partial \ln\mu$
   \cite{Aronov:95}, which in particular implies a sign change at
   optimal doping, \emph{i.e.} near the ETT.

In order to understand the nonmonotonic dependence of the critical
   temperature $T_c$ as a function of lattice strain, we argue that a
   strain-induced deformation of the lattice varies the parameters in
   Eq.~(\ref{eq:dispersion}), so that $T_c$ attains its optimal value
   close to the ETT.
To this aim, one has to recognize that applied pressure (hydrostatic
   pressure or anisotropic stress) can in principle 
   modify all the parameters in the model, so that several
   contributions to the overall pressure dependence of $T_c$ can be
   identified \cite{Angilella:96}.
In particular, pressure is expected to modify the overall hole doping
   level $\delta$.
Moreover, an intrinsic source of variation for $T_c$
   is expected to come from the dependence of the hopping
   parameters as well as of the coupling constant on the lattice
   spacings.
In Ref.~\cite{Angilella:96}, a phenomenological dependence of the
   hopping parameters as well as the coupling constant on hydrostatic
   pressure has been assumed \cite{Angilella:96}.
Here, we restrict to the case $\delta = \mbox{const}$, and argue
   that the main parameter driving the deformation of the Fermi
   line in the case of in-plane epitaxial strain be the hopping ratio
   $r$.

\section{Numerical results and discussion}
\label{sec:numerical}

A standard mean-field approximation of Eq.~(\ref{eq:hamiltonian})
   yields the BCS gap equation \cite{Fehrenbacher:95,Angilella:96}:
\begin{equation}
\Delta_\bk = -\frac{1}{N} \sum_{\bk^\prime} V_{\bk\bk^\prime}
   \chi_{\bk^\prime} \Delta_{\bk^\prime} ,
\label{eq:gap}
\end{equation}
where $\Delta_\bk \equiv \Delta g_\bk$ is the gap energy, $\chi_\bk =
   (2E_\bk )^{-1} \tanh (\frac{1}{2}\beta E_\bk )$ the pair
   susceptibility, 
   $E_\bk = \sqrt{\xi_\bk^2 + \Delta_\bk^2}$ the upper branch of the
   superconducting excitation spectrum, and $\beta = (k_{\mathrm B}
   T)^{-1}$ the inverse temperature.
Eq.~(\ref{eq:gap}) must be supplemented by the equation defining the
   band filling $n$, or equivalently the hole doping $\delta=1-n$,
\begin{equation}
n = 1 - 2 \sum_\bk \xi_\bk \chi_\bk .
\label{eq:delta}
\end{equation}

At $T=T_c$, $\Delta \to0$, and Eq.~(\ref{eq:gap}) can be linearized as
\begin{equation}
1 + \lambda \frac{1}{N} \sum_\bk g^2_\bk \chi^c_\bk = 0,
\label{eq:BCSlin}
\end{equation}
where $\chi_\bk^c = (2\xi_\bk )^{-1} \tanh (\beta_c \xi_\bk /2 )$.
Eqs.~(\ref{eq:BCSlin}) and (\ref{eq:delta}) can be solved
   self-consistently for the critical temperature $T_c$ and the
   chemical potential $\mu$, at fixed hole content $\delta$, for a
   given hopping ratio $r$.

\begin{figure}[t]
\centering
\includegraphics[height=0.9\columnwidth,angle=-90]{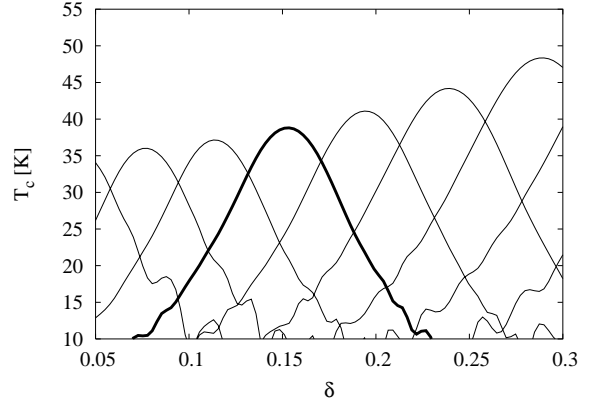}
\caption{%
Critical temperature $T_c$ as a function of hole doping $\delta$,
   Eq.~(\protect\ref{eq:delta}), for fixed hopping
   ratio $r=t^\prime /t = 0\div 0.5$ ($t=0.4$~eV,
   $\lambda = -0.45$~eV).
Along each curve, one recovers the typical bell-shaped dependence of
   $T_c$ on $\delta$, the maximum occurring close to the ETT.
The thicker line corresponds to $r=0.182$, for which $T_c$ attains a
   maximum of $\approx 40$~K for $\delta\approx 0.15$, as observed
   experimentally for LSCO.
}
\label{fig:mesh1}
\end{figure}

\begin{figure}[t]
\centering
\includegraphics[height=0.9\columnwidth,angle=-90]{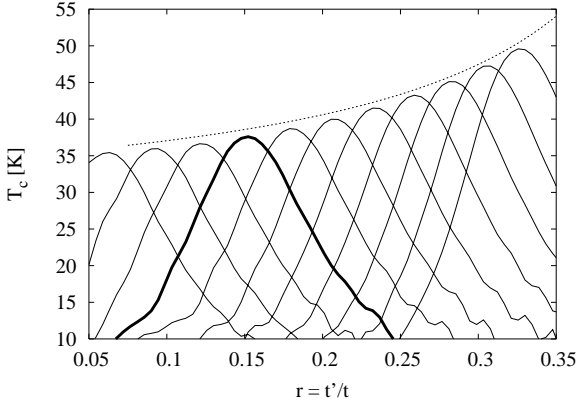}
\caption{%
Critical temperature $T_c$ as a function of hopping ratio $r=t^\prime
   /t$ (other parameters as in Fig.~\protect\ref{fig:mesh1}).
Along each curve, $\delta = \mathrm{const}$ ($\delta = 0.05\div 0.3$,
   as in Fig.~\protect\ref{fig:mesh1}; the thicker line corresponds to 
   $\delta=0.125$).
One recognizes a nonmonotonic
   dependence of $T_c$ on strain, as observed experimentally, the
   maximum in $T_c$ being attained close to the ETT. 
One also recovers the direct correlation between $T_c^{\mathrm{max}}$
   on each curve and the hopping ratio $r$ (dashed line, vertically
   shifted, for clarity)
   \protect\cite{Pavarini:01,Angilella:01,Angilella:02}.
}
\label{fig:mesh2}
\end{figure}

Figs.~\ref{fig:mesh1} and \ref{fig:mesh2} show our numerical results
   for $T_c$ as a function of $\delta$, for fixed
   values of the hopping ratio $r$ in the meaningful range
   $0\div 0.5$, and for $T_c$ as a function of $r$, for fixed
   hole content $\delta$, respectively ($t=0.4$~eV,
   $\lambda = -0.45$~eV, yielding an optimal $T_c \approx 40$~K at
   $\delta \approx 0.15$ for $r\approx0.2$, as observed
   experimentally in LSCO).

\begin{figure}[t]
\centering
\includegraphics[bb=76 66 546 511,clip,width=0.47\columnwidth]{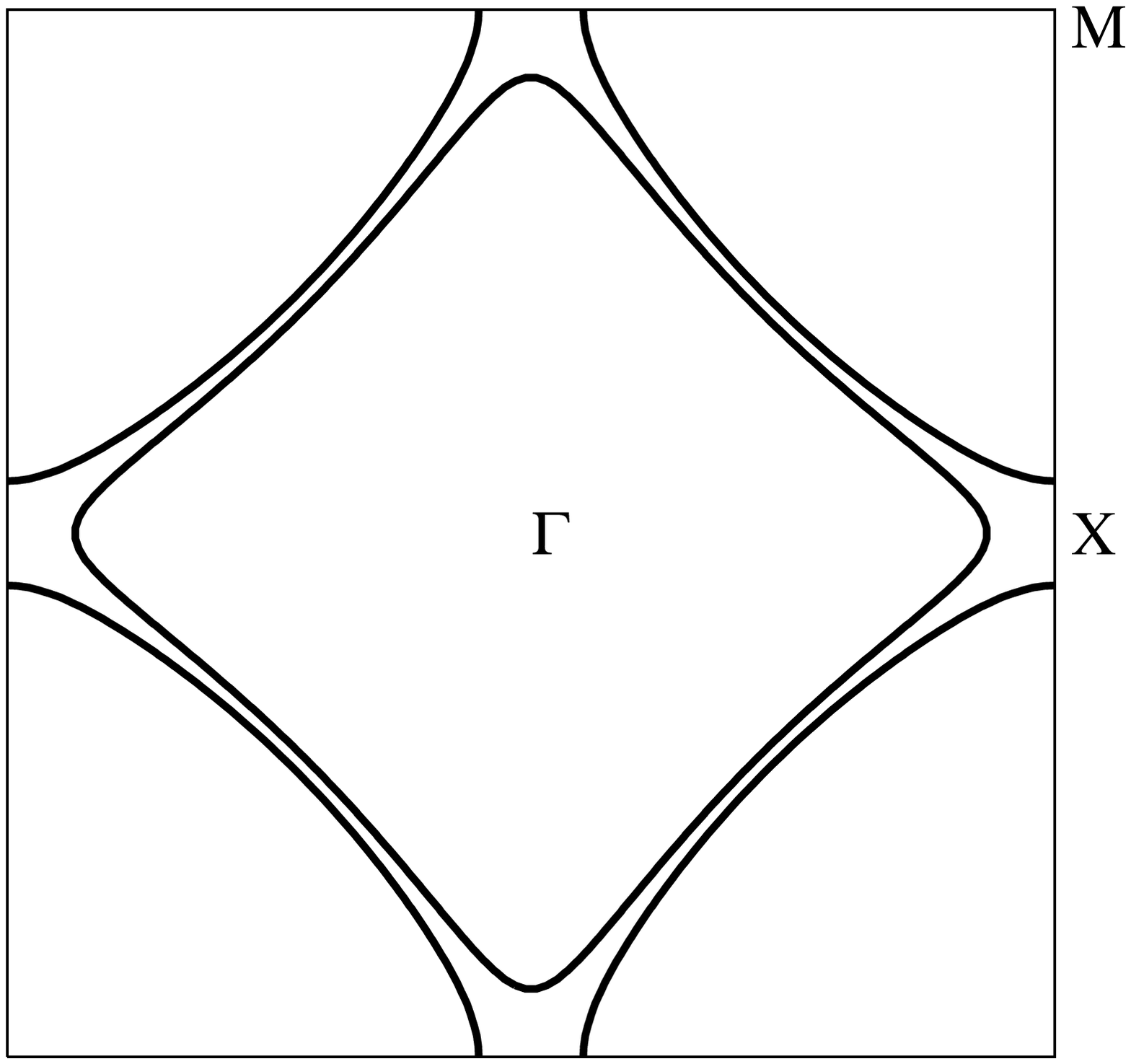}
\includegraphics[bb=76 66 546 511,clip,width=0.47\columnwidth]{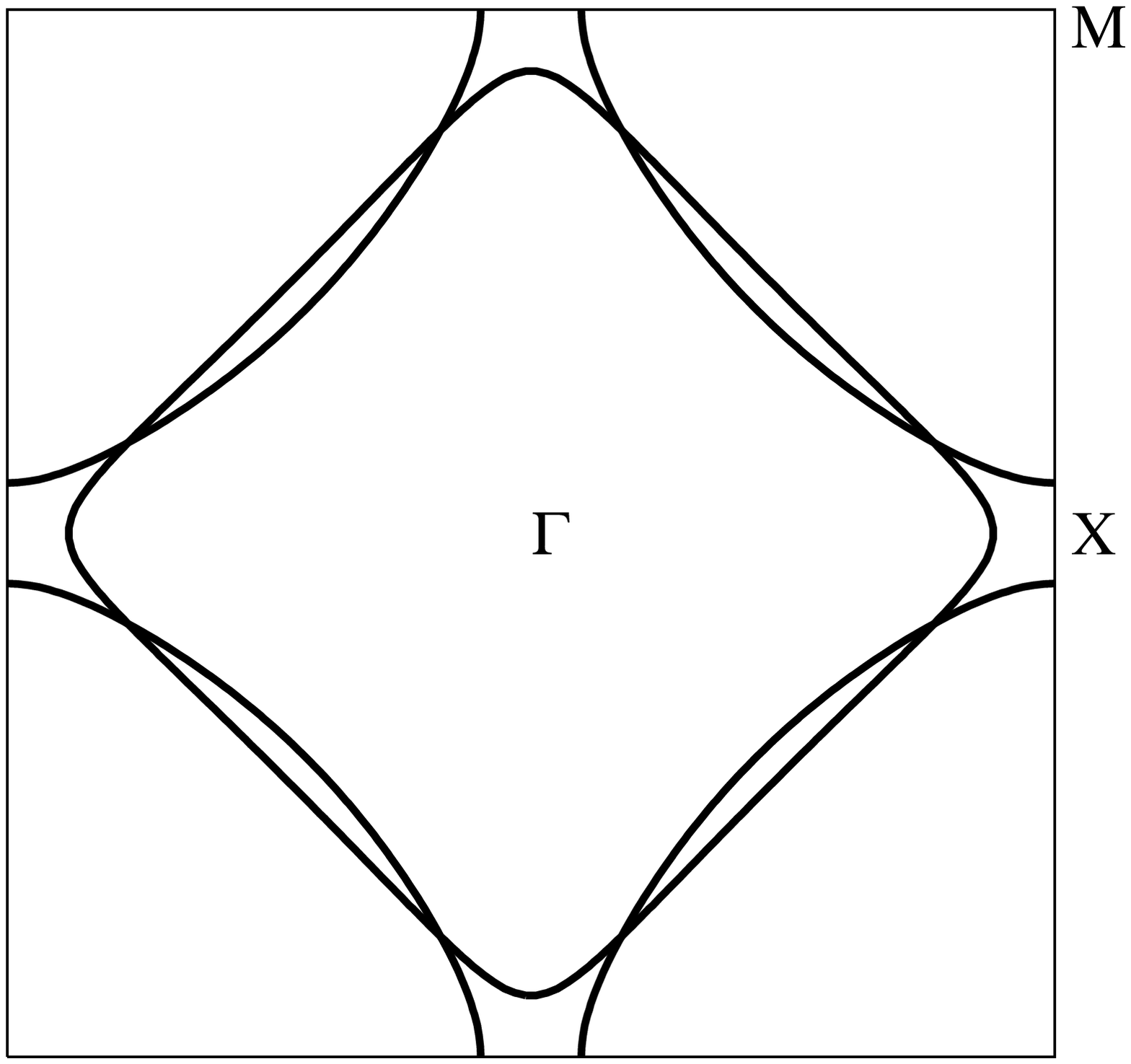}
\caption{%
Typical Fermi lines $\xi_\bk = 0$, Eq.~(\protect\ref{eq:dispersion}),
   at either side of an ETT.
{\sl Left:} The ETT is driven by a variation of the hole content
   $\delta$, at constant $r$ ($r=0.182$, corresponding to the thicker
   plot in Fig.~\protect\ref{fig:mesh1}).
{\sl Right:} The ETT is driven by a (strain-induced) variation of the
   hopping ratio 
   $r$, at constant $\delta$ ($\delta=0.125$, thicker plot in
   Fig.~\protect\ref{fig:mesh2}).
In both cases, the Fermi line changes topology, from a hole-like
   contour, centered around the $M=(\pi,\pi)$ point, to an
   electron-like contour, centered around the $\Gamma$ point.
}
\label{fig:ETT}
\end{figure}

In Fig.~\ref{fig:mesh1}, each curve corresponds to a given
   band dispersion relation, Eq.~(\protect\ref{eq:dispersion}), fixed
   by a constant value of the hopping ratio $r$. 
The topology of the Fermi line $\xi_\bk = 0$ evolves from a hole-like
   to an electron-like contour as $\delta$ increases ($\mu$
   decreases), as depicted in Fig.~\ref{fig:ETT} (left).
Here, the ETT is driven by a variation of the hole content $\delta$,
   which in turn implies a change in chemical potential $\mu$, through
   Eq.~(\ref{eq:delta}). 
For a given band structure ($r=\mathrm{const}$), one recognizes the
   typical bell-shaped dependence of $T_c$ on doping, as observed
   experimentally \cite{Zhang:93,Wijngaarden:99}.
A maximum in $T_c$ is found close, though not exactly at, the ETT
   \cite{Angilella:01,Angilella:02}.

On the other hand, epitaxial strain in thin films may realize the
   conditions assumed in Fig.~\ref{fig:mesh2}, \emph{viz.} a
   modification of the lattice spacings induce a variation of the
   hopping parameters in Eq.~(\ref{eq:dispersion}), and therefore in
   the hopping ratio $r$.
Here, we assume that we can neglect the strain dependence of the
   NN hopping parameter $t$, mainly fixing the scale
   for $T_c$, compared to that of the NNN hopping parameter
   $t^\prime$, whose value determines the actual shape of the Fermi
   line at the ETT.
Indeed, within the extended H\"uckel theory \cite{Hoffmann:63}, the
   hopping parameters $t$ an $t^\prime$ can be roughly approximated by 
   the overlap integrals between NN Cu $3d_{x^2 - y^2}$ and O $2p_x$
   orbitals, and NNN O $2p_x$ and O $2p_y$ orbitals, respectively (see 
   Fig.~1 in Ref.~\cite{Pavarini:01}).
Due to the weaker overlap of the latter two orbitals, it is to be
   expected that for moderate strain $t^\prime$ increases much faster
   than $t$ as the CuO$_2$ unit cell is compressed, provided that the
   tetragonal symmetry of the lattice is preserved \cite{Angilella:96}.
We can also neglect the strain dependence of the hole
   content $\delta$, which in the case 
   of LSCO close to optimal doping is known to be weakly dependent on
   hydrostatic pressure \cite{Murayama:91}.
On the other hand, our main conclusions should not be affected by a
   strain-dependent coupling constant $\lambda$ (see, however,
   Ref.~\cite{Angilella:96}).

Fig.~\ref{fig:mesh2} displays our numerical results for $T_c$ as a
   function of the hopping ratio $r=t^\prime /t$.
Each curve corresponds to a constant value of the hole doping
   $\delta$.
Assuming an approximately linear dependence of the hopping ratio $r$
   on the in-plain microscopic strain $\varepsilon_{aa} =
   \varepsilon_{bb}$, and neglecting the strain dependence of all other
   parameters, one recovers a nonmonotonic dependence of $T_c$ on
   strain, as observed experimentally, in agreement with the
   phenomenological model of Sec.~\ref{sec:phenomenological}.
In particular, the maximum in $T_c$ is attained close to the ETT.
At variance with the case considered in Fig.~\ref{fig:mesh1}, the
   topology change of the Fermi line at fixed hole doping is here
   driven by a strain-induced variation of the band parameters
   (Fig.~\ref{fig:ETT}, right).
One also recovers the direct correlation between the critical
   temperature at optimal doping for each curve, $T_c^{\mathrm{max}}$,
   and the hopping ratio $r$ (Fig.~\ref{fig:mesh2}, dashed line), as
   analytically 
   found in Ref.~\protect\cite{Angilella:01,Angilella:02}, and
   observed experimentally for many high-$T_c$ cuprates
   \protect\cite{Pavarini:01}.

\section{Conclusions}
\label{sec:conclusions}

Within a Ginzburg-Landau approach, we have proposed a
   phenomenological model for the critical temperature $T_c$ as a
   function of microscopic strain $\boldsymbol{\varepsilon}$ in a
   tetragonal cuprate superconductor.
For small strains, the model predicts an increase of
   $T_c$ when the size of the CuO octahedron increases.
Such a behavior is rather generic to high-$T_c$ superconductors, as
   has been very recently confirmed by the observed increase of $T_c$
   up to 117~K in lattice-expanded doped fullerites \cite{Schoen:01}.

On the other hand, for fixed in-plane or apical strains, $T_c$
   displays a nonmonotonic behavior on either apical or in-plain
   microstrains, respectively.
Such a behavior is recovered also for the dependence of $T_c$ on
   hydrostatic pressure $P$, in agreement with the $T_c$ vs. $P$ plots of
   most cuprate compounds \cite{Wijngaarden:99}.
Moreover, under epitaxial strain, we find a monotonically decreasing
   $T_c$ in the experimentally accessible range $-0.006\leq
   \varepsilon_{\mathrm{epi}} \leq 0.006$, with a sharp maximum just
   below the lower bound of the mentioned range, in good qualitative
   and quantitative agreement with the experimental results for
   epitaxially strained LSCO \cite{Locquet:98,Locquet:98a}.

From a microscopic point of view, a nonmonotonic strain
   dependence of the critical temperature in the high-$T_c$ cuprates
   has been interpreted as due to the proximity to  
   an electronic topological transition (ETT).
The quasi-2D band structure generic to cuprates implies a
   topology change of the Fermi line $\xi_\bk = 0$, evolving from a
   hole-like to an electron-like contour in the 1BZ.
A variation (in the most general sense of variational calculus) of the
   band-dispersion $\xi_\bk$
   induces a change in various observable properties, such as
   $T_c$ (thus, a \emph{functional} of $\xi_\bk$), with a maximum
   occurring at, or close to, the ETT. 

Assuming a tight-binding parametrization of the band structure, one
   way to `vary' $\xi_\bk$ is that of changing the hole content.
Physically, this is what is most commonly realized by doping in the
   experiments, and what has been usually considered in the theoretical
   literature \cite{Onufrieva:99a,Onufrieva:99b}.
Such a situation corresponds to the assumption of a \emph{rigid} band,
   namely an electronic band whose structure does not depend on its
   filling.

Here, we considered another class of `variations' of $\xi_\bk$,
   \emph{i.e.} a change in the band parameters, at fixed hole content.
A modification of the in-plane band parameters can be induced by in-plane
   epitaxial strain, through a change in the lattice spacings, without
   perturbing the tetragonal symmetry of the lattice.
The idea of an ETT driven by a modification of the band structure, at
   fixed hole content, as contrasted to an ETT induced by doping or
   hydrostatic pressure for a rigid band, is particularly relevant
   for LSCO, where the hole content is known to be practically
   independent of pressure \cite{Murayama:91}.

Thus, a numerical analysis of a minimal microscopic model for $d$-wave
   superconductivity close to an ETT allowed us to recover a
   nonmonotonic dependence 
   of $T_c$ on the hopping ratio $r$, measuring the distortion of the
   electronic band under in-plane epitaxial strain.
At constant doping, a variation of $r$ modifies the topology of the
   Fermi line and drives an electronic topological transition, with
   $T_c$ attaining the maximum value close to the ETT.
Such a result enables us to justify, from a microscopic point of
   view, the proposed phenomenological model for $T_c$ as a function
   of microstrain in the cuprates.

\begin{acknowledgement}

G.G.N.A. acknowledges partial support from the E.U. through the
   F.S.E. Program.
P.C. and P.P.G. acknowledge partial support from the M.U.R.S.T. through
   Progetto Cofinanziato 2000 ``Modelli Matematici per la Scienza dei
   Materiali''; from the E.U. through the TMR Contract
   FMRX-CT98-0229 ``Phase Transitions in Crystalline Solids''; and
   from the GNFM (INDAM), Contract ``Effetti della deformazione sulla
   temperatura critica dei superconduttori.''

\end{acknowledgement}

\bibliographystyle{mprsty}
\bibliography{Angilella,a,b,c,d,e,f,g,h,i,j,k,l,m,n,o,p,q,r,s,t,u,v,w,x,y,z,zzproceedings}

\end{document}